\documentclass[prl,twocolumn]{revtex4}
\usepackage{graphicx}
\usepackage{amssymb,amsmath,bm}
\usepackage{enumerate}

\newcommand{\be}{\begin{eqnarray}}
\newcommand{\ee}{\end{eqnarray}}

\begin{document}

\title{Polymer additives in fluid turbulence and distributed chaos }

\author{A. Bershadskii}

\affiliation{
ICAR, P.O. Box 31155, Jerusalem 91000, Israel
}

\begin{abstract}

The fluids and polymers have different fundamental symmetries. Namely, the Lagrangian relabeling symmetry of fluids is absent for polymers (while the translational and rotational symmetries are still present). This fact results in spontaneous breaking of the relabeling symmetry in fluid turbulence even at a tiny polymer addition. Since helicity conservation in inviscid fluid motions is a consequence of the relabeling symmetry (due to the Noether's theorem) violation of this conservation by the polymer additives results in the strong effects in the distributed chaos. The distributed chaos in turbulence with the spontaneously broken relabeling symmetry is characterized by stretched exponential spectra $\propto \exp(-k/k_{\beta})^{\beta}$ with $\beta =2/5$. The spectral range of this distributed chaos is extended in direction of the small wavenumbers and $k_{\beta}$ becomes much larger in comparison with the pure fluid (Newtonian) case. This results in substantial suppression of small-scale turbulence and large-scale mixing enhancement. Good agreement with results of direct numerical simulations (DNS) and experimental data has been established for a channel flow (DNS), for Rayleigh-Taylor turbulent convection (DNS) and for Rayleigh-Benard turbulent convection (experiment).

\end{abstract}

\maketitle

\section{Introduction}

The very strong influence of dilute polymers solutions on turbulent flows indicates an underlying spontaneous symmetry breaking phenomenon. What fundamental symmetry can be spontaneously 
broken by this additive? The answer is about obvious. It is well known that in elasticity the location of the particle in an unstressed configuration is often used as the material coordinate. For fluid particles the situation is drastically different because these particles can be rearranged without the creation of stress. The absence of an unique unstressed reference configuration of a fluid is called particle relabeling symmetry (or merely relabeling symmetry, since this symmetry represents the arbitrariness of providing  fluid elements with
Lagrangian labels - any transformations that conserve the
density of labels \cite{salmon}). The standard elasticity has no such fundamental symmetry,  while the other fundamental symmetry groups: rotations and translations, are still present \cite{mas}. Therefore, even small additives of the polymers to fluids can result in spontaneous breaking of the fundamental relabeling symmetry. In order to understand main consequences of such spontaneous symmetry breaking one should recall that the fundamental symmetries, due to the Noether's theorems, result in the fundamental conservation laws. Namely, the translational symmetry (homogeneity) in time and space results in conservation of energy and momentum (Birkhoff-Saffman invariant \cite{saf},\cite{dav}), respectively. The rotational symmetry (isotropy) results in the angular momentum conservation (Loitsyanskii invariant \cite{my}). The relabeling symmetry results in conservation of helicity in inviscid fluids  \cite{y},\cite{pm},\cite{fs},\cite{mor} (in the fluids with non-zero viscosity the reflexional symmetry should be broken in order to "switch on" the integral viscosity dissipation/production of helicity). Therefore, the spontaneous breaking of the relabeling symmetry should be strongly related to violation of the helicity invariance (even in the viscous turbulence). Of course, this is valid for the {\it dilute} polymer solutions only, because for a strong solution one cannot already speak about a spontaneous symmetry breaking.

\section{Spontaneous breaking of relabeling symmetry}

Spontaneous breaking of a fundamental symmetry can strongly affect distributed chaos in turbulence (see, for instance, Ref. \cite{b1}). Spectrum of the distributed chaos in turbulence is a weighted superposition of exponentials converged into a stretch exponential
$$
E(k ) \simeq \int_0^{\infty} \mathcal{P} (\kappa)~ e^{-(k/\kappa)}  d\kappa  \propto \exp-(k/k_{\beta})^{\beta}  \eqno{(1)}
$$
The parameter $\beta$ is related to scaling exponent $\alpha$ of the group velocity of the waves (pulses) driving the chaos
$$
\upsilon (\kappa ) \sim \kappa^{\alpha}     \eqno{(2)}
$$
Namely,
$$
\beta =\frac{2\alpha}{1+2\alpha}   \eqno{(3)}
$$ 
 In isotropic homogeneous turbulence the momentum correlation integral (the Birkhoff-Saffman invariant, a consequence of the space translational symmetry, see Introduction)
$$   
I_2 = \int  \langle {\bf u} ({\bf x},t) \cdot  {\bf u} ({\bf x} + {\bf r},t) \rangle d{\bf r}  \eqno{(4)}
$$ 
dominates the scaling Eq. (2) and provides $\alpha = 3/2$ from the dimensional considerations: 
$$
\upsilon (\kappa ) \propto ~I_2^{1/2}~\kappa^{3/2} \eqno{(5)}
$$
Corresponding $\beta =3/4$ (Eq. (3)).
 
   Spontaneous breaking of the space translational symmetry (homogeneity) replaces the Birkhoff-Saffman invariant by vorticity correlation integral
$$
\gamma = \int_{V} \langle {\boldsymbol \omega} ({\bf x},t) \cdot  {\boldsymbol \omega} ({\bf x} + {\bf r},t) \rangle_{V}  d{\bf r} \eqno{(6)}
$$   
and, due to the dimensional considerations
$$
\upsilon (\kappa ) \propto |\gamma|^{1/2}~\kappa^{1/2} \eqno{(7)}
$$
provides $\beta =1/2$. 

   If one wants to use the same approach as in the Ref. \cite{b1} for the case of spontaneous breaking of the relabeling symmetry, then a natural choice of the invariant (instead of the Birkhoff-Saffman invariant) is the helicity correlation integral (see Introduction)
$$   
I = \int  \langle  h ({\bf x},t) \cdot   h ({\bf x} + {\bf r},t) \rangle d{\bf r}  \eqno{(8)}
$$   
(the Levich-Tsinober invariant \cite{lt},\cite{fl},\cite{l}). However, it can be readily shown that this choice results in $\alpha=\beta=0$. 

  Fortunately, there is another helicity based invariant \cite{mor}
$$
\mathcal{H} = \int  h ({\bf r},t) d{\bf r}  \eqno{(9)}
$$
with
$$
\frac{d\mathcal{H}}{dt}=\mathrm{Vi} + \mathrm{El} \eqno{(10)}
$$  
where $\mathrm{Vi}$ and $\mathrm{El}$ are the viscous and elastic terms, respectively. Then from the dimensional considerations
$$
\upsilon (\kappa ) \propto |\mathrm{El}|^{1/3}~\kappa^{1/3} \eqno{(11)}
$$
i.e. $\alpha = 1/3$ and from Eq. (3) $\beta = 2/5$ for the spontaneously broken relabeling symmetry. It can be shown that $k_{\beta} \propto 1/|El|$, that is important for suppression of small-scale turbulence.

  It should be noted that the spontaneous symmetry breaking can have a localized character \cite{bt} - in blobs. 
    
\section{Turbulent viscoelastic channel flows}

In recent direct numerical simulations (DNS) \cite{tgm},\cite{tmg} of fully developed
turbulent channel flows the viscoelastic fluid was modeled using the FENE-P constitutive
equation, with a ratio of the Newtonian viscosity to the total
zero-shear viscosity of $\beta_0 = 0.9$ and at the highest friction Reynolds number available to date, $Re_{\tau 0} = 1000$. Two different regimes of drag reduction were considered: a flow with medium percentage drag reduction (30\%) and one with a high percentage of drag reduction (58\%), corresponding polymer lengths $L = 30$ and $L = 100$, respectively (friction Weissenberg numbers $We_{\tau 0} = 50$ and $We_{\tau 0} = 115$).

   Figure 1 shows power spectral density of the streamwise velocity at the wall-normal position $y^+ = 99$ for Newtonian fluid in log-log scales. The data are taken from Ref. \cite{tgm}. The solid curve is drawn in order to indicate 
the stretched exponential spectrum Eq. (1) with $\beta =1/2$ corresponding to the distributed chaos with spontaneously broken space translational symmetry Eq. (7) (cf Ref. \cite{b1}).  
   
\begin{figure}
\begin{center}
\includegraphics[width=8cm \vspace{-1.2cm}]{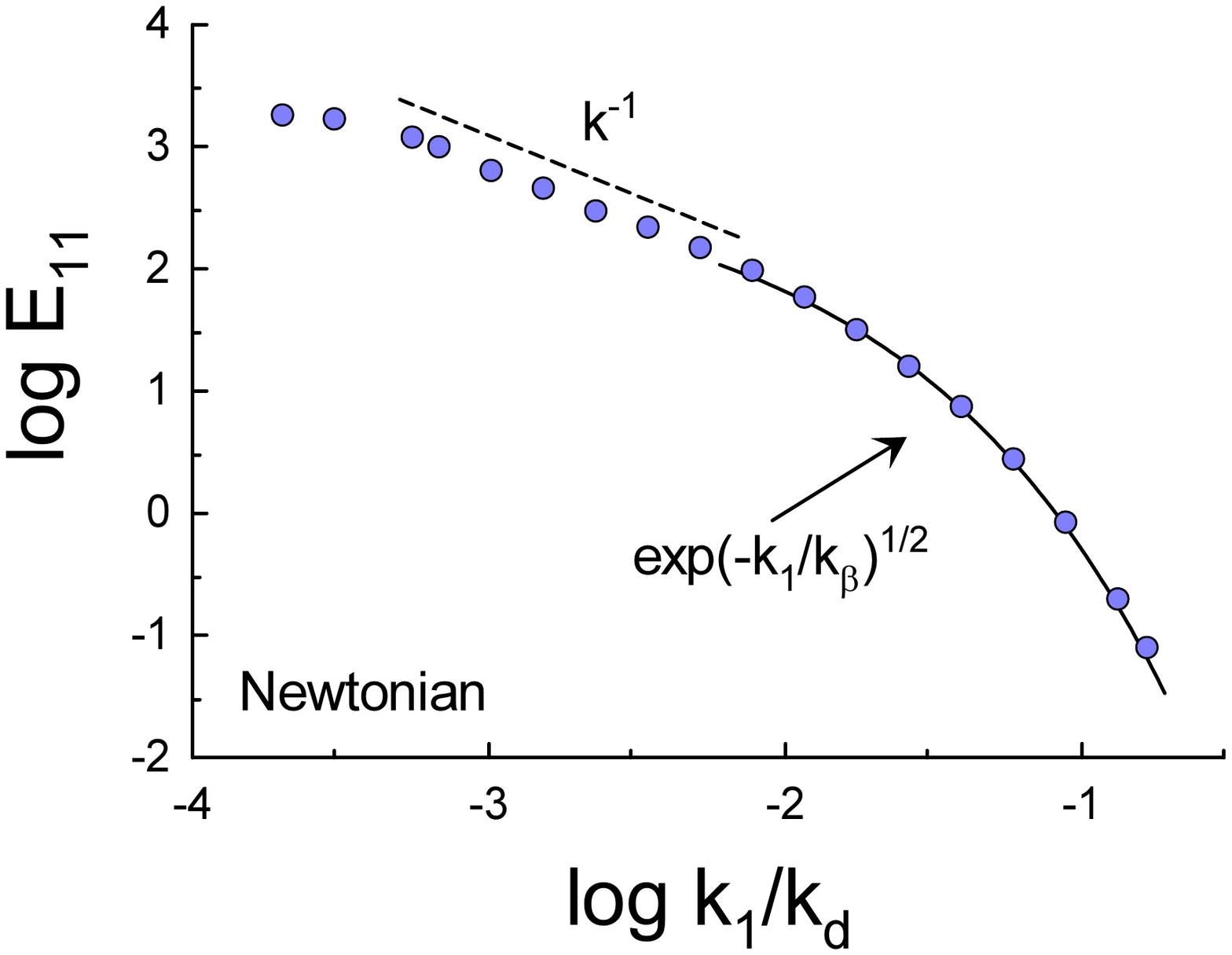}\vspace{-3.5cm}
\caption{\label{fig1} Power spectrum of the streamwise velocity at the wall-normal position $y^+ = 99$ for Newtonian fluid in log-log scales. $k_d$ is the dissipative (Kolmogorov) scale.}
\end{center}
\end{figure}
\begin{figure}
\begin{center}
\includegraphics[width=8cm \vspace{-1.38cm}]{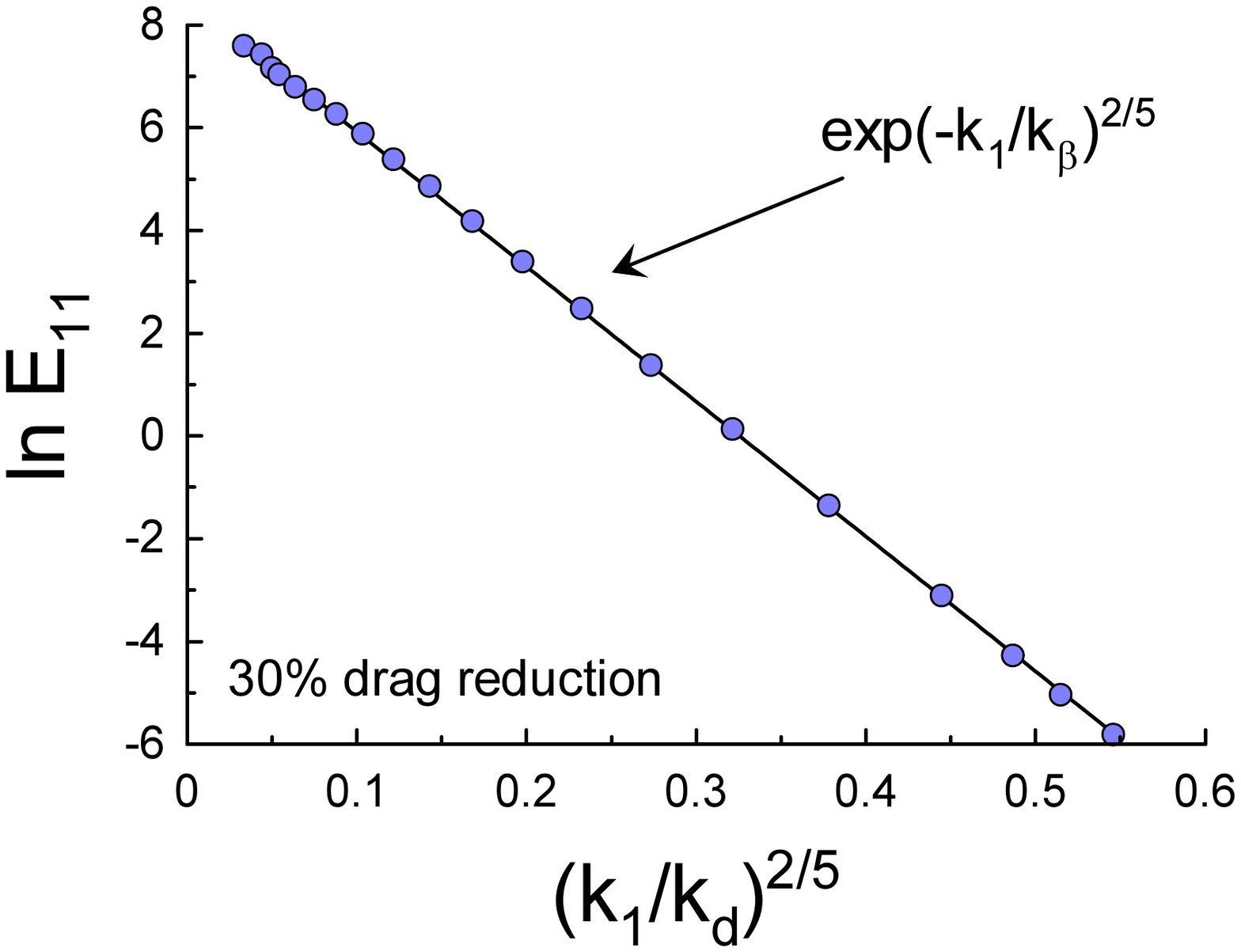}\vspace{-3.8cm}
\caption{\label{fig2} The same as in Fig. 1 but with the polymer additive (the 30\% drag reduction). The straight line indicates the stretched exponential decay Eq. (1) with $\beta =2/5$ - spontaneous breaking of the relabeling symmetry.} 
\end{center}
\end{figure}

Figure 2 shows the same flow as in Fig. 1 but with the polymer additive (the 30\% drag reduction). The scales in the figure are chosen in order to indicate the stretched exponential 
Eq. (1) with $\beta =2/5$ Eq. (11) (the spontaneous breaking of the relabeling symmetry) as a straight line. 

  Figure 3 shows the same flow as in Fig. 2 but at the {\it mid-channel center plane}. The scales in the figure are chosen in order to indicate the stretched exponential 
Eq. (1) with $\beta =1/2$ corresponding to the distributed chaos with spontaneously broken space translational symmetry Eq. (7) (cf Fig.1 for Newtonian fluid and Fig. 2 for $y^+ =99$). One can see that at the mid-channel center plane (unlike at $y^+= 99$) the streamwise velocity power spectrum indicates the same distributed chaos as for Newtonian fluid, despite the 30\% drag reduction. 

  However, the 58\% drag reduction affects the flow even at the mid-channel center plane. Figure 4 shows the same flow as in Fig. 3 but with the polymer additive corresponding to the 58\% drag reduction. The scales in the figure are chosen in order to indicate the stretched exponential 
Eq. (1) with $\beta =2/5$ Eq. (11) (the spontaneous breaking of the relabeling symmetry) as a straight line (cf Fig. 2). 

\begin{figure}
\begin{center}
\includegraphics[width=8cm \vspace{-1cm}]{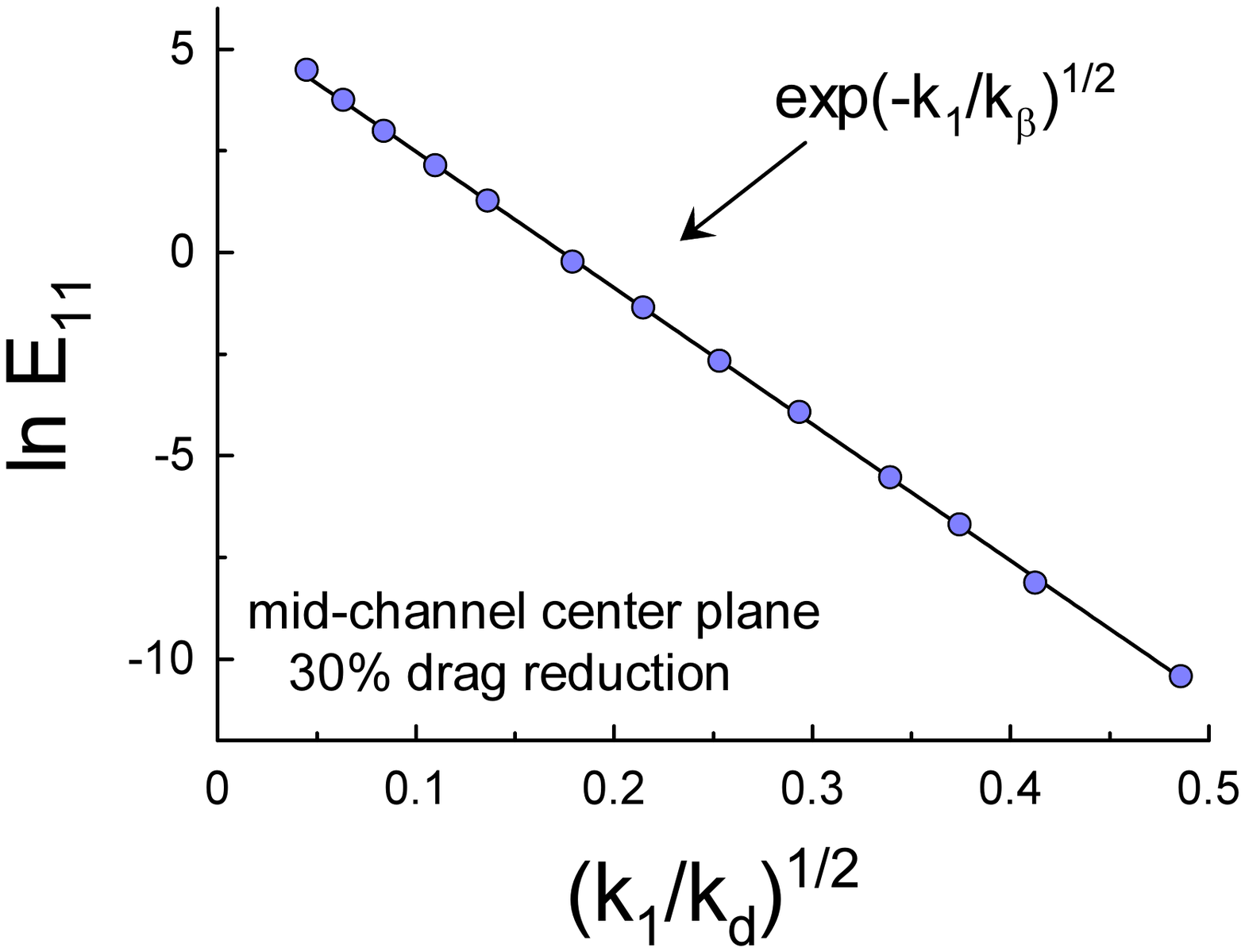}\vspace{-3.7cm}
\caption{\label{fig3} The same as in Fig. 2 but at the  mid-channel center plane. The straight line indicates the stretched exponential decay Eq. (1) with $\beta =1/2$ - spontaneous breaking of the translational symmetry.}
\end{center}
\end{figure}
\begin{figure}
\begin{center}
\includegraphics[width=8cm \vspace{-1.5cm}]{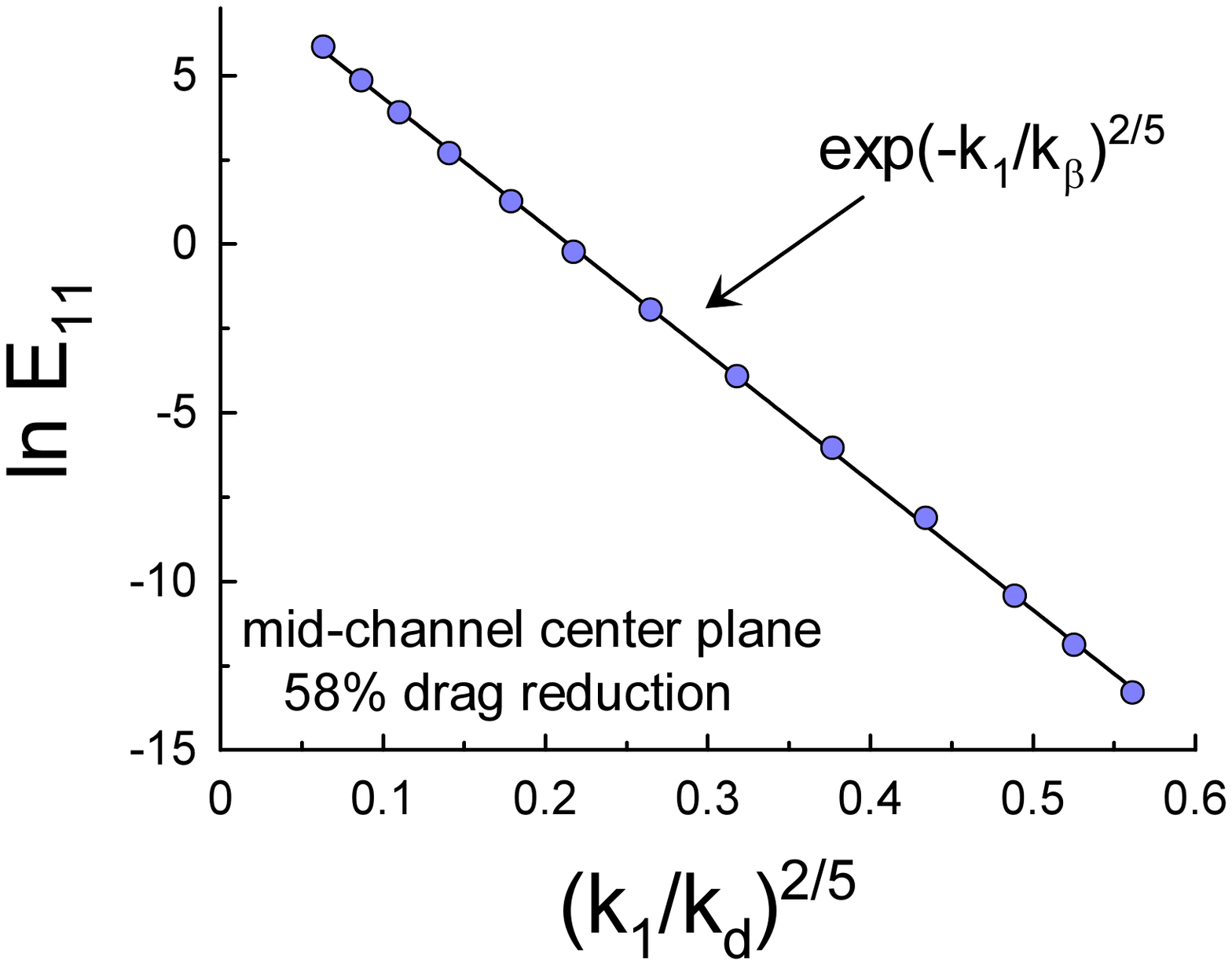}\vspace{-3.35cm}
\caption{\label{fig4} The same as in Fig. 3 but for 58\% drag reduction. The straight line indicates the stretched exponential decay Eq. (1) with $\beta =2/5$ - spontaneous breaking of the relabeling symmetry.} 
\end{center}
\end{figure}

\begin{figure}
\begin{center}
\includegraphics[width=8cm \vspace{-0.83cm}]{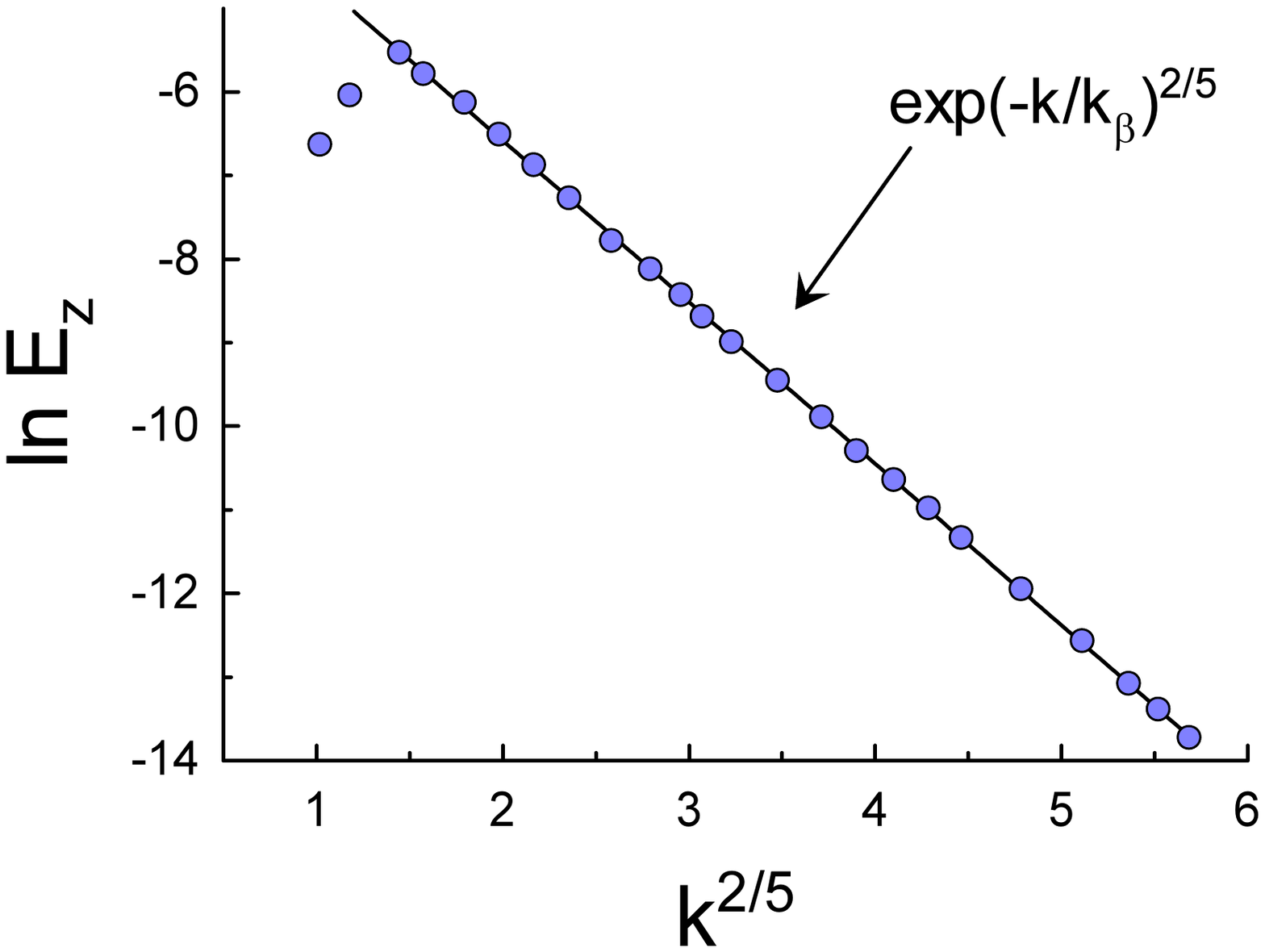}\vspace{-4cm}
\caption{\label{fig5} Power spectrum of the vertical component of velocity field for the Zimm relaxation time $\tau_p = 1$ and at computational time equal to $3.1\tau$.}
\end{center}
\end{figure}

\section{Polymer additives in Rayleigh-Taylor turbulent convection}

In previous section we have considered effects of the polymer additives in the {\it bulk} of the channel flow, i.e. outside of the near-wall region with its complex large-scale coherent structures. In  a recent DNS reported in Ref. \cite{bmm} the Rayleigh-Taylor turbulent convection with a tiny amount of long-chain polymers additives were simulated using a generalization
of the Boussinesq approximation to a viscoelastic case with the Oldroyd-B model (at Prandtl number $Pr=1$, low Atwood number - $A$, and dilute polymers
solution). It is shown that a drag reduction also occurs in the complete absence of walls and results in enhancement of mixing at large scales, in suppression of small-scale fluctuations and, as a consequence, in a considerable enhancement of heat transport. 
  The computational characteristic time $\tau=(L_z/Ag)^{1/2}$ ($L_z$ is the vertical side of the computational domain, and $g$ - is gravity acceleration). 

   Figure 5 shows power spectrum of the vertical component of velocity field for the Zimm relaxation time \cite{bird} $\tau_p = 1$ and at computational time equal to $3.1\tau$. The data were taken from Ref. \cite{bmm}. The solid straight line indicates the stretched exponential decay Eq. (1) with $\beta = 2/5$ Eq. (11), i.e. the spontaneous breaking of the relabeling symmetry (cf Figs. 2 and 4). Figure 6 shows analogous power spectrum for the Zimm relaxation time $\tau_p = 2$. 

\begin{figure}
\begin{center}
\includegraphics[width=8cm \vspace{-1.35cm}]{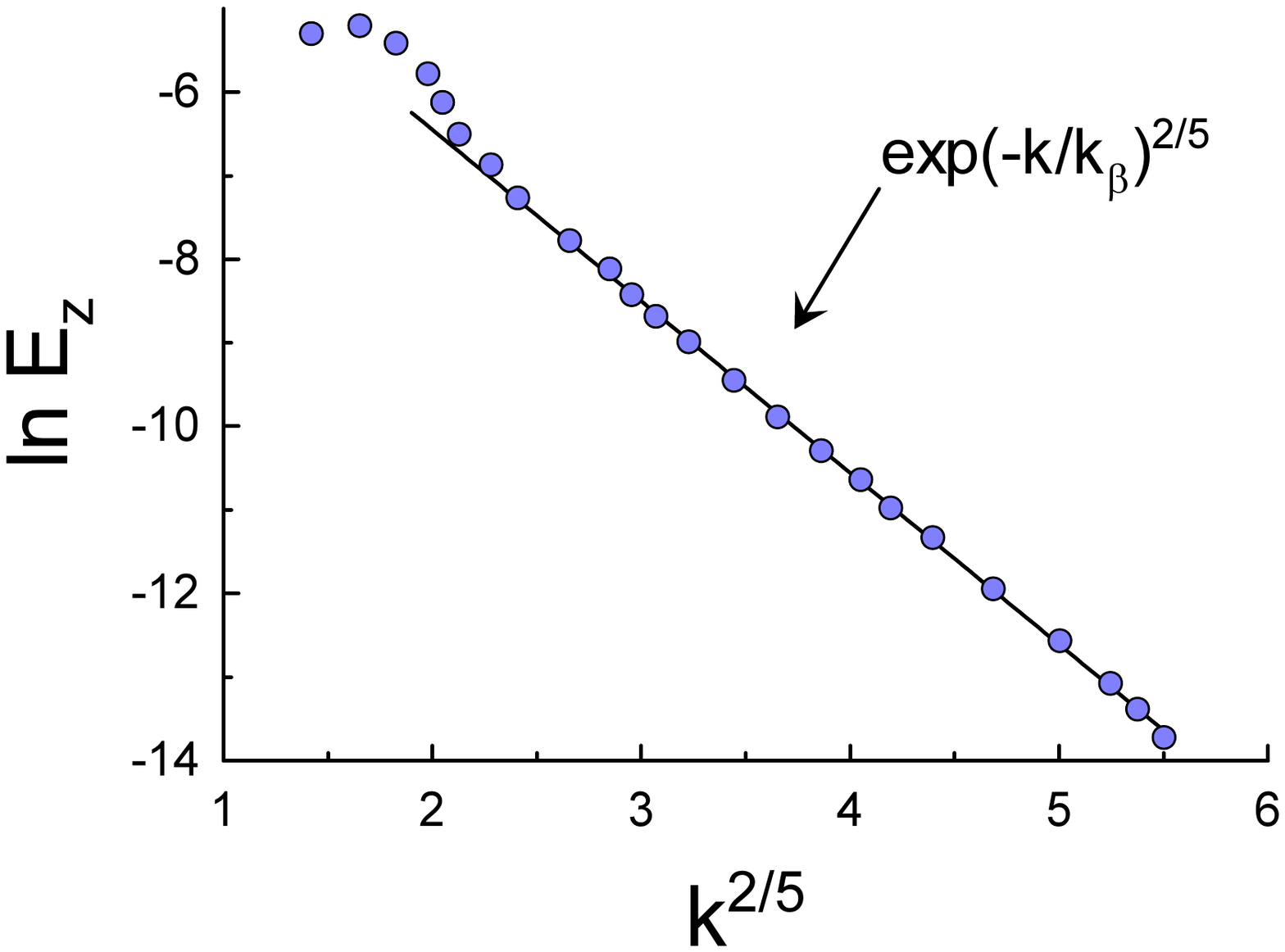}\vspace{-3.8cm}
\caption{\label{fig6} The same as in Fig. 5 but for the Zimm relaxation time $\tau_p = 2$}
\end{center}
\end{figure}

\section{Polymer additives in the bulk of turbulent Rayleigh-Benard convection}

 \begin{figure}
\begin{center}
\includegraphics[width=8cm \vspace{-1.3cm}]{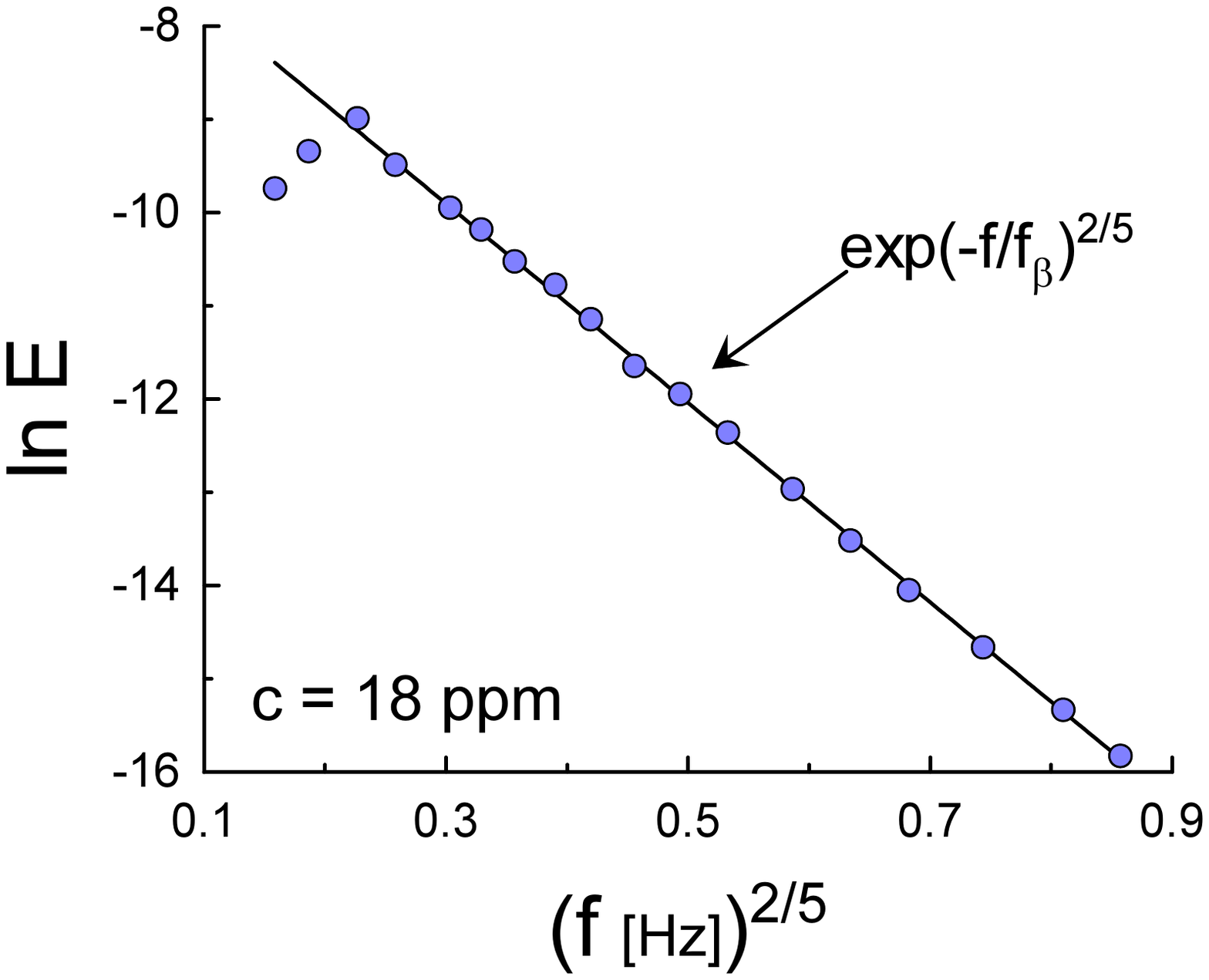}\vspace{-3.8cm}
\caption{\label{fig7} Velocity power spectrum (data are taken from the Ref. \cite{xx}) at 
the polymer concentration $c=18$ ppm. }
\end{center}
\end{figure}
In turbulent Rayleigh-Benard convection (unlike the Rayleigh-Taylor convection) there are two distinct flow regions: thermal boundary layers near walls and a bulk region. These 
two regions react to the polymer additives in rather different ways. As in the case of the channel flow we will be interested in the bulk region here. In a recent experiment, reported in Ref. \cite{xx}, effects of tiny polymer additives (Polyacrylimid in water) on the bulk region were studied in an upright cylindrical convection cell with diameter D=19.3 cm and height
H=19.5 cm. For the Newtonian case (without the polymer additives) Rayleigh number $Ra = 6.18 \times 10^9$. The Weissenberg number $We =\tau_p/\tau_{\eta}= 0.1$, where $\tau_{\eta}$ is the Kolmogorov's characteristic time. The velocity measurements were made in the center of the convection cell using two dimensional laser Doppler velocimetery. 

    Figure 7 shows velocity power spectrum (data are taken from the Ref. \cite{xx}) at 
the polymer concentration $c=18$ ppm (parts per million by weight). The solid straight line indicates the stretched exponential decay Eq. (1) with $\beta = 2/5$ Eq. (11), i.e. the spontaneous breaking of the relabeling symmetry (cf Figs. 2 and 4-6).\\ 

  It should be noted that for considered above turbulent flows {\it without} the polymer additives 
$k_{\beta} \sim k_e$, where $k_e$ is an estimate of the most energy containing wavenumber. With the polymer additives the situation is drastically different: $k_{\beta} \ll k_e$. This difference results in exponential like Eq. (1) suppression of the small-scale turbulence for the case of the spontaneous breaking of the relabeling symmetry (Section II). This suppression alone can lead to the observed coherent large-scale mixing enhancement. This effect is also considerably strengthened by the fact that the range of the distributed chaos is extended into large-scale direction (smaller wavenumbers $k$) with the spontaneous breaking of the relabeling symmetry.


\begin{thebibliography}{99}
\bibitem{salmon} R. Salmon, Ann. Rev. Fluid Mech, {\bf 20}, 225 (1988).
\bibitem{mas} J. E. Marsden et. al., arXiv:math/0005034 (2000).
\bibitem{y} A. Yahalom, arXiv:solv-int/9407001 (1994); J. Math. Phys. {\bf 36} 1324 (1995).
\bibitem{pm} N. Padhye and P. J. Morrison,  Phys. Lett. A {\bf 219}, 287 (1996).
\bibitem{fs} Y. Fukumotoa , H. Sakumab, Procedia IUTAM {\bf 7} 213 ( 2013) .
\bibitem{mor} J. J. Moreau was first to discover the helicity conservation: J. J. Moreau, C. R. Akad. Sci. Paris {\bf 252}, 2810, (1961) and first to relate it to a symmetry group (V. I. Arnold, J. M´ec., {\bf 5}, 19, 1966) of fluid element labeling via the Noether theorem: J.J. Moreau, Seminaire D’analyse Convexe, Montpellier 1977 , Expose no: 7 (1977). H. K. Moffatt pointed out the topological difference between fluids and polymers related to the helicity conservation in fluids: H. K. Moffatt, J . Fluid Mech. {\bf 106}, 27 (1981).
\bibitem{saf} P. G. Saffman, J. Fluid. Mech. {\bf 27}, 551 (1967).
\bibitem{dav} P. A. Davidson P.A. Turbulence in rotating, stratified and electrically conducting fluids. (Cambridge University Press, 2013).
\bibitem{my} A. S. Monin, A. M. Yaglom, Statistical Fluid Mechanics, Vol. II, (Dover Pub. NY, 2007).
\bibitem{b1} A. Bershadskii, arXiv:1601.07364 (2016).
\bibitem{lt} E. Levich and A. Tsinober, Phys. Lett. A {\bf 93}, 293 (1983).
\bibitem{fl} A. Frenkel and E. Levich, Phys. Lett. A {\bf 98}, 25 (1983).
\bibitem{l} E. Levich, Concepts of Physics {\bf VI}, 239 (2009).
\bibitem{bt} A. Bershadskii and A. Tsinober, Phys. Rev. E {\bf 48}, 282 (1993).
\bibitem{tgm} http://www.tsfp-conference.org/proceedings/2013/v2/\\con1e.pdf
\bibitem{tmg} L. Thais, G. Mompean,  and T.B. Gatski, Journal of Non-Newtonian Fluid Mechanics {\bf 200}, 165 (2013).
\bibitem{bmm} G. Boffetta, A. Mazzino, and S. Musacchio, Phys. Rev. E {\bf 83}, 056318 (2011).
\bibitem{bird} R. B. Bird et al. Dynamics of Polymeric Liquids (Wiley Interscience, New York,
1987).
\bibitem{xx} Y-C. Xie et al., arXiv:1511.03822; J. Fluid Mech.{\bf 784}, R3 (2015).






\end{thebibliography}
\end{document}